\documentclass{aastex} \usepackage{emulateapj5}
\newcommand{\msun}{M$_{\odot}$}
\newcommand{\asecdot}[2]{\mbox{#1$\stackrel {\prime \prime}{_{\bf \cdot}}$#2}}
\newcommand{\adegdot}[2]{\mbox{#1$\stackrel {\circ}{_{\bf \cdot}}$#2}}
\newcommand{\kms}{km\,s$^{-1}$}       % km/s

\slugcomment{Accepted by ApJL. Last changed 2001-10-30 (AB)}

\shorttitle{Atomic gas in the rotating $\beta$ Pic disk}
\shortauthors{Olofsson et al.}

\begin{document}

\title{Widespread atomic gas emission reveals the rotation\\ of the $\beta$\,Pictoris disk\footnotemark}

\author{G\"{o}ran Olofsson, Ren\'e Liseau and
 Alexis Brandeker}
\affil{Stockholm Observatory, SCFAB, SE-106 91 Stockholm, Sweden}
\email{olofsson@astro.su.se}

\footnotetext[1]{Based on observations collected at the European Southern Observatory, Chile}
%\renewcommand{\thefootnote}{\fnsymbol{footnote}}
%\footnotetext[1]{Based on observations collected at the European Southern
%Observatory, Chile}
%\renewcommand{\thefootnote}{\arabic{footnote}}

\begin{abstract}

We present high resolution \ion{Na}{1}~D spectroscopy of the $\beta$\,Pic disk, and
the resonantly scattered sodium emission can be traced from less than 30\,AU to
at least 140\,AU from the central star. This atomic gas is co-existent with the dust
particles, suggestive of a common origin or source. The disk rotates towards us
in the south-west and away from us in the north-east. The velocity pattern
of the gas finally provides direct evidence that the faint linear
feature seen in images of the star is a circumstellar disk in Keplerian
rotation. From modelling the spatial distribution of the \ion{Na}{1} line profiles
we determine the effective dynamical mass to be ($1.40\pm0.05$)\,\msun, which is 
smaller than the stellar mass, 1.75\,\msun. We ascribe this difference to
the gravity opposing radiation pressure in the \ion{Na}{1} lines. We argue that
this is consistent with the fact that Na is nearly completely ionised
throughout the disk (\ion{Na}{1}/Na $< 10^{-4}$). The total column density of
sodium gas is $N$(Na)=$10^{15}\,{\rm cm}^{-2}$.
\end{abstract}


%\objectname{beta Pic}

\section{Introduction}

The $\beta$\,Pic disk has been the subject of intense studies ever since it
was shown to have excess emission in the thermal infrared by the IRAS
mission \citep{aum85}. The dust particles giving rise to the infrared
emission also scatter the stellar light and coronographic techniques
revealed an elongated structure \citep{smi84} that recently
has been extremely well characterised by means of HST observations
\citep{hea00}. Although the general impression of these images is a high
degree of smoothness and symmetry, \citet{hea00} show that the peak
of the elongated scattered emission has wiggles that suggests that the
inner part of the disk is tilted relative to the outer part. There is
also evidence for asymmetries and structures in the radial light
distribution of this inner disk. Asymmetries have also been observed in
the mid-IR \citep{pan97}.

These asymmetries and the remarkable discovery \citep{vid94,lag00}
of comet-like bodies occasionally occulting
the star and giving rise to high velocity absorption lines reveal
dynamical activity that generally is interpreted as evidence for
planetary disturbances \citep{art97,aug01}.

When it comes to the gas component in the disk, the situation is at
present confusing. Absorption spectroscopy in the optical and UV
shows a stable component at the same radial velocity as the star
\citep{vid86,lag98}. One would expect
radial expansion of this gas, caused by the radiation pressure of the
star and as a possible explanation a ring of atomic hydrogen has been
proposed \citep{lag98}. If present, such a ring must be close
to the star as widespread \ion{H}{1} is not detected at 21\,cm \citep{fre95}.
CO has been detected by means of UV spectroscopy \citep{rob00}
but sensitive searches for CO emission in the radio region have
failed to detect this molecule \citep{lis98}. Molecular
hydrogen is not detected in UV absorption spectra \citep{lec01}
but on the other hand, ISO/SWS spectroscopy \citep{thi01}
indicates H$_2$ emission lines that, if real (the S/N is low),
require large amounts of molecular hydrogen in a spatial distribution
that has a void in the line of sight towards the star.

So far it has not been possible to directly probe the velocity pattern
along the disk, lacking the detection of spatially resolved gas emission
(except a possible detection of \ion{Fe}{2} emission very close to the star,
\citet{lec00}). In the present paper we report the
detection of the resonance \ion{Na}{1} doublet lines at 5990\,\AA\ and 5996\,\AA.

\section{Observations and results}

$\beta$\,Pic was observed on January 2-4, 2001, with the Echelle
spectrograph EMMI on the 3.5\,m NTT (New Technology Telescope) at ESO-La
Silla, Chile. For the \ion{Na}{1} D1/D2 observations, we used a 5\,arcmin long
slit and therefore omitted the cross-disperser. Instead, we inserted an
order-sorting interference filter, which was centred on 5890\,\AA\ and 60\,\AA\ 
wide. We have measured the filter profile and are in control of the
leakage from the neighbouring orders (less than 3\% and 4\% respectively).
With the wavelength resolution of $6\cdot 10^{4}$ ($\Delta v$ = 5\,\kms) this
instrumental set-up can be viewed as some sort of a `spectroscopic
coronograph', where the high dispersion is used to prevent the
saturation of the detector (CCD) by the very bright stellar point
source. Still, the overheads of the observing time were dominated by
detector readout times, since integration times of individual
observations were typically only 300\,s. The total integration time was
chosen such that the co-added spectra should permit the detection of the
faint light scattered off the disk by the dust, which was also achieved.
Parallel with the disk (at position angle \adegdot{30}{75}, \citet{kal95}),
this time was 2.5\,hr,
whereas we spent in total 36\,min for observations with the slit oriented
perpendicular to the disk. The seeing values determined from these
observations are \asecdot{1}{2}, whereas the width of the slit was kept at
\asecdot{1}{0} in order to sample (on the CCD) the spectral resolution at
the Nyquist frequency.

\begin{figure}
\plotone{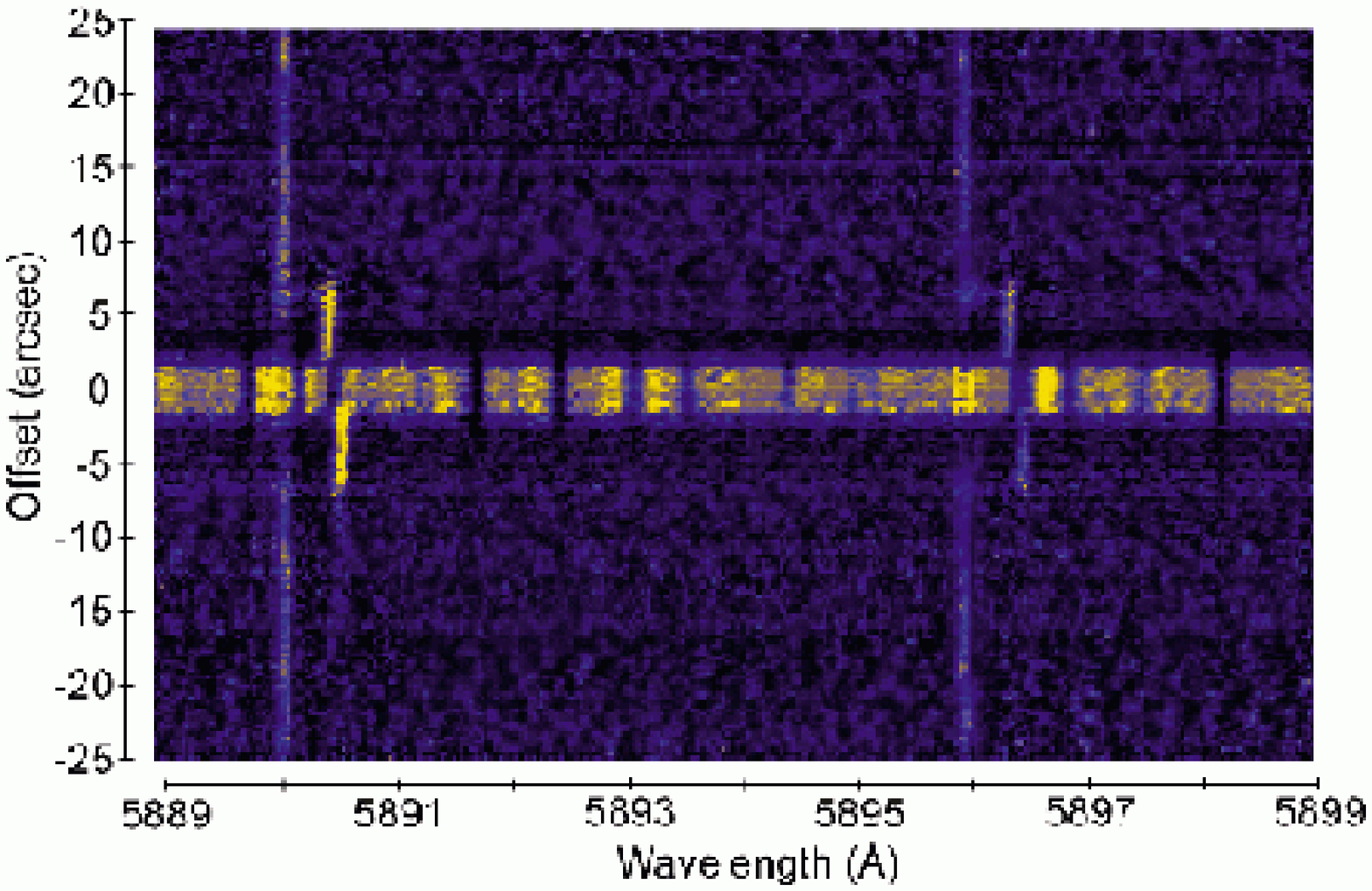}
\caption{The spectral region around the \ion{Na}{1} D lines toward 
$\beta$\,Pictoris.
The dispersion is along the horizontal axis, with wavelengths increasing
to the right, and the scale is 0.035\,\AA\ per pixel (1.8\,\kms ${\rm pxl}^{-1}$).
The spatial dimension is along the vertical axis, with the 1\arcsec\ wide and
300\arcsec\ long slit oriented along the position angle \adegdot{30}{75}. The scale is
0.27\,arcsec per pixel (5\,AU ${\rm pxl}^{-1}$), with positive values toward the
south-west and negative values toward the north-east. The (deliberately
very much reduced) stellar continuum at the centre of the figure is
covered with telluric absorption lines, whereas the terrestrial
ionospheric \ion{Na}{1} D1/D2 emission lines extend over the entire vertical
space. The \ion{Na}{1} line emission from the $\beta$\,Pic disk is seen redshifted
with respect to the sky lines and with the intensity ratio 1:2. On
either side of the star, the disk lines are either blue- or redshifted
with respect to the systemic rest wavelength, represented by the disk
gas absorption seen against the stellar continuum. \label{fig1}}
\end{figure}

\begin{figure}
\plotone{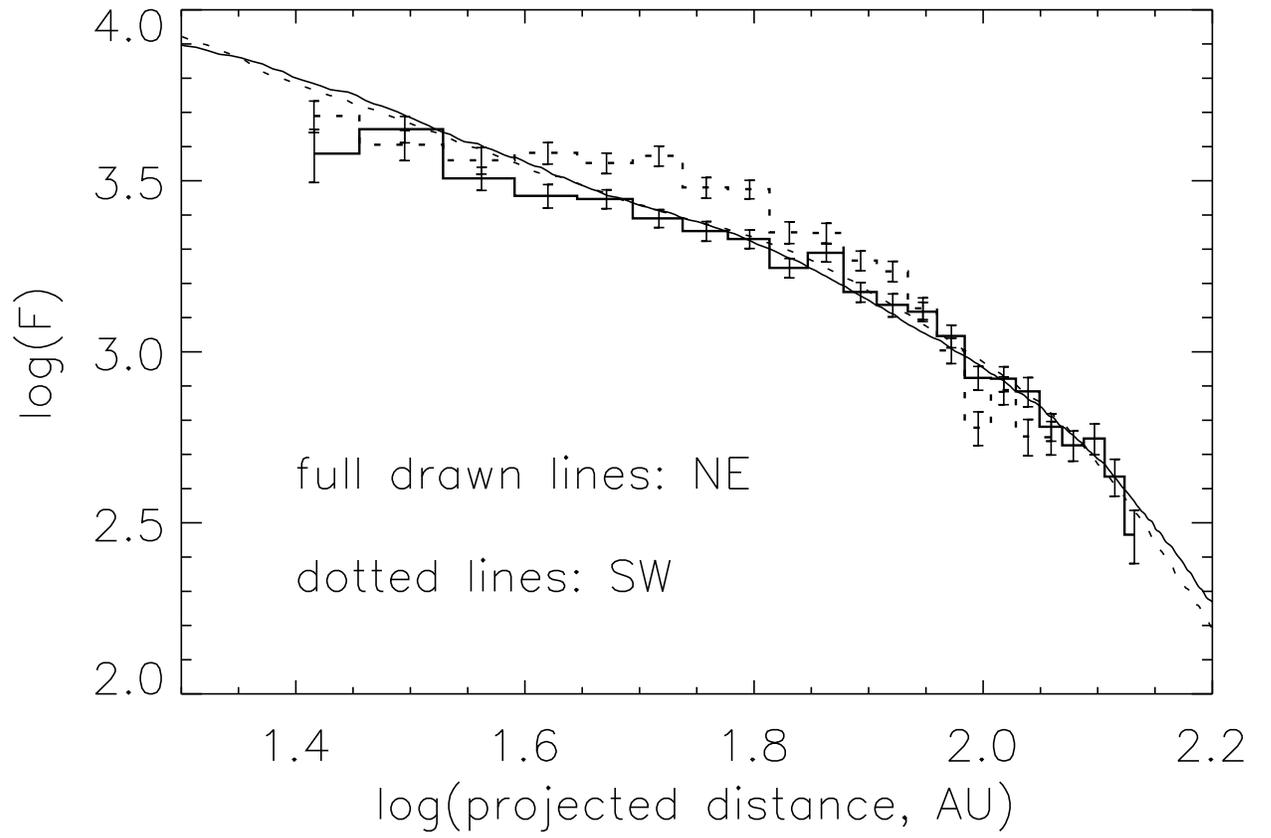}
\caption{ The radial distribution, for the two sides of the disk, of the
\ion{Na}{1} D2 line emission
with error bars compared to that of the scattered stellar radiation caused
by the dust (from \citet{hea00}). The flux scale is arbitrary and adjusted
to facilitate the comparison.\label{fig2}}
\end{figure}

In addition to the two telluric emission lines,
the parallel observations also show spatially extended \ion{Na}{1}~D emission,
both lines consistently redshifted to the heliocentric velocity of
$\beta$\,Pic (+21\,\kms). These observations are displayed in Fig.~\ref{fig1}.
In Fig.~\ref{fig2}, the distribution of the \ion{Na}{1} emission along the slit,
on either side of the stellar spectrum, is compared to the corresponding
distribution of the dust. Evidently, both gas and dust show similar light
distributions, suggestive of the co-existence of these species in the $\beta$\,Pic
disk. 

\section{Discussion}

\subsection{The rotation of the disk}

\begin{figure}
\plotone{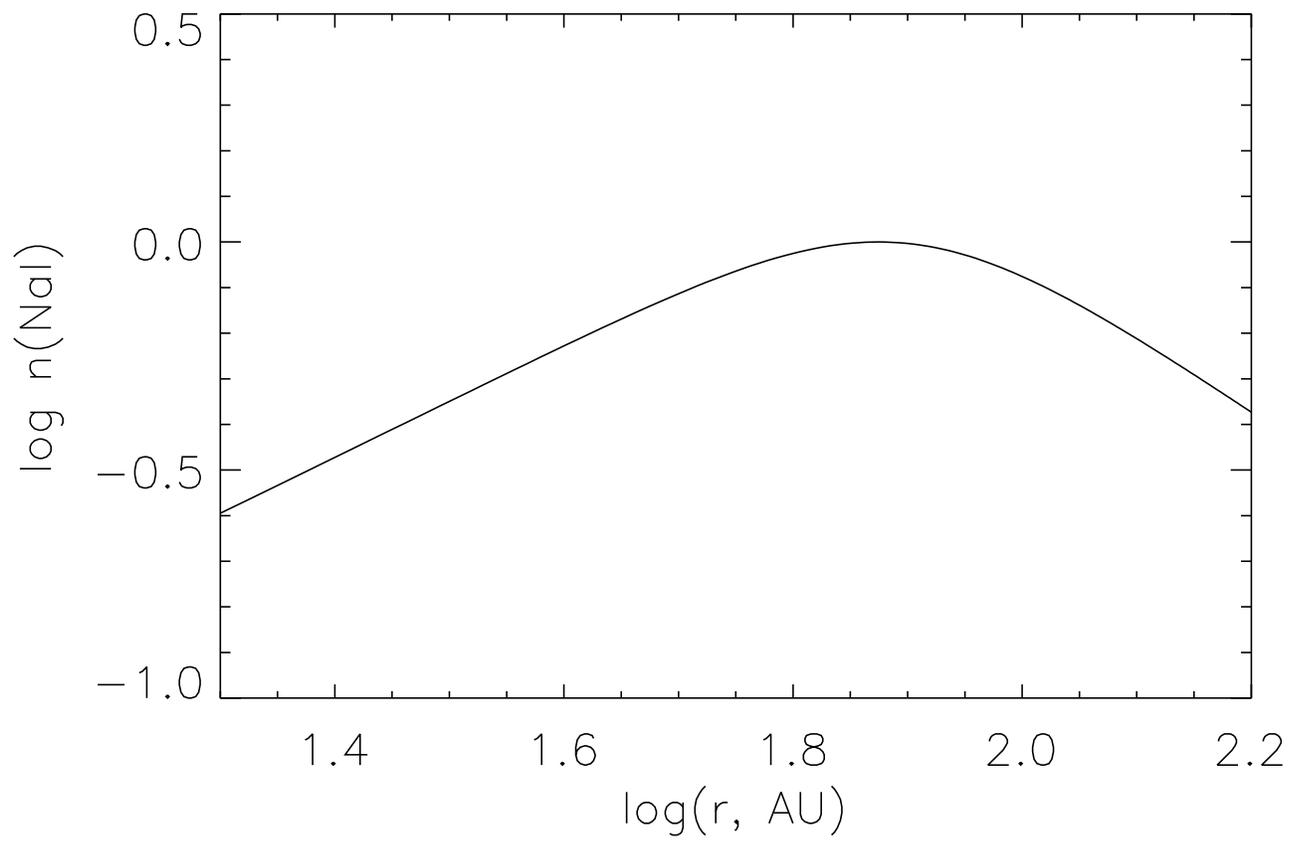}
\caption{The derived radial distribution for \ion{Na}{1} (normalised at the
peak).\label{fig3}}
\end{figure}

\begin{figure}
\plotone{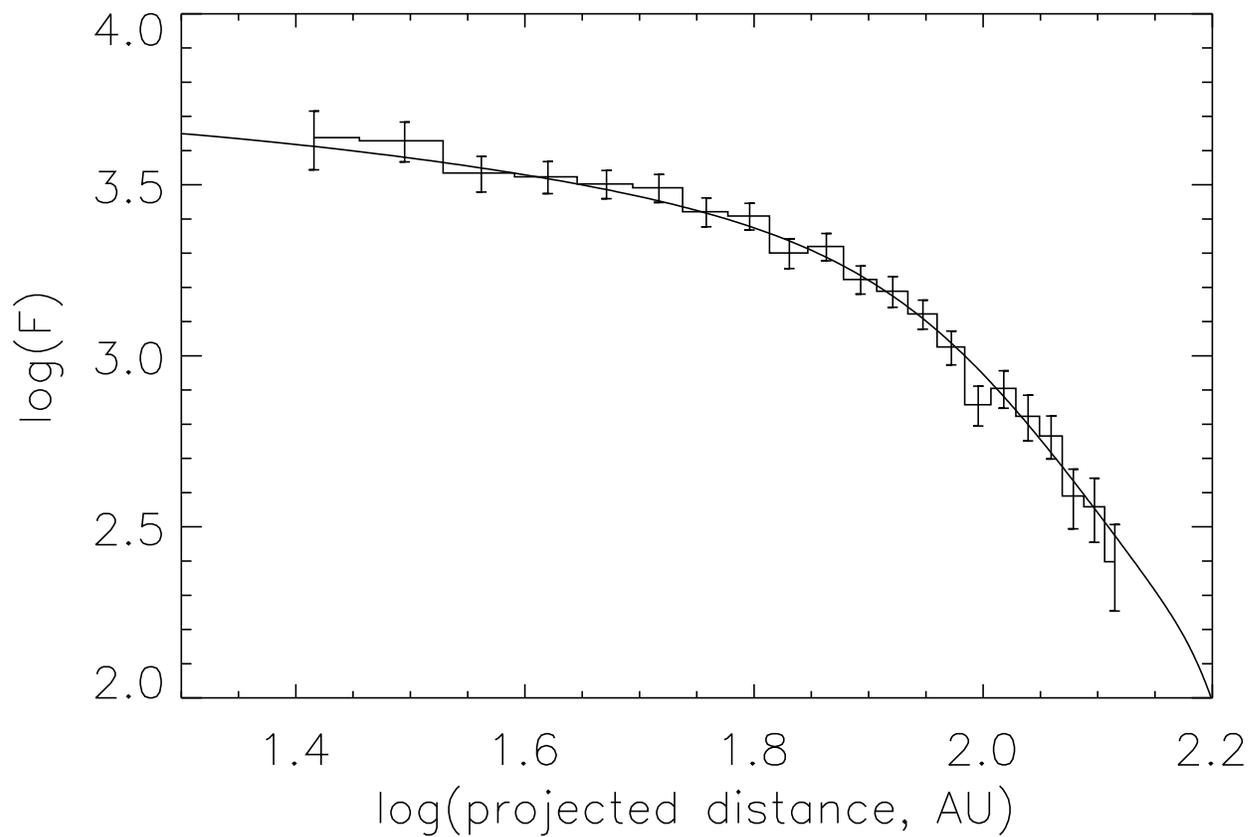}
\caption{The projected radial distribution of the \ion{Na}{1} D2 emission of the
model (smooth line) compared to the observed (average of the SW and NE
sides).\label{fig4}}
\end{figure}

We first address the question of Keplerian rotation. Justified by the
indications that there is no significant outward motion, we simply
assume circular rotation. The distance to the star is 19.3\,pc \citep{cri97}.
In order to interpret the velocity pattern, we must first
find a radial density distribution of the Na gas component that results
in the observed light distribution. For the moment, we ignore the
difference between the two sides (Fig.~\ref{fig2}) and use the average light
distribution. Adopting a functional relation that is suitable for a broken power law:
\begin{displaymath}
n({\rm Na\,I}) \propto \left [(r/a)^{2b}+(r/a)^{2c}\right ]^{-\frac{1}{2}}
\end{displaymath}
where $n$(\ion{Na}{1}) is the volume density of \ion{Na}{1} at the radial
distance $r$ (within the 1\arcsec\ slit), we get a quite satisfactory fit
to the observed projected light distribution with the following numerical
values: $a=89.5$\,AU, $b=-1.28$ and $c=3.44$. The distribution function is
shown in Fig.~\ref{fig3} and the fit to the observations in Fig.~\ref{fig4}.

\begin{figure}
\plotone{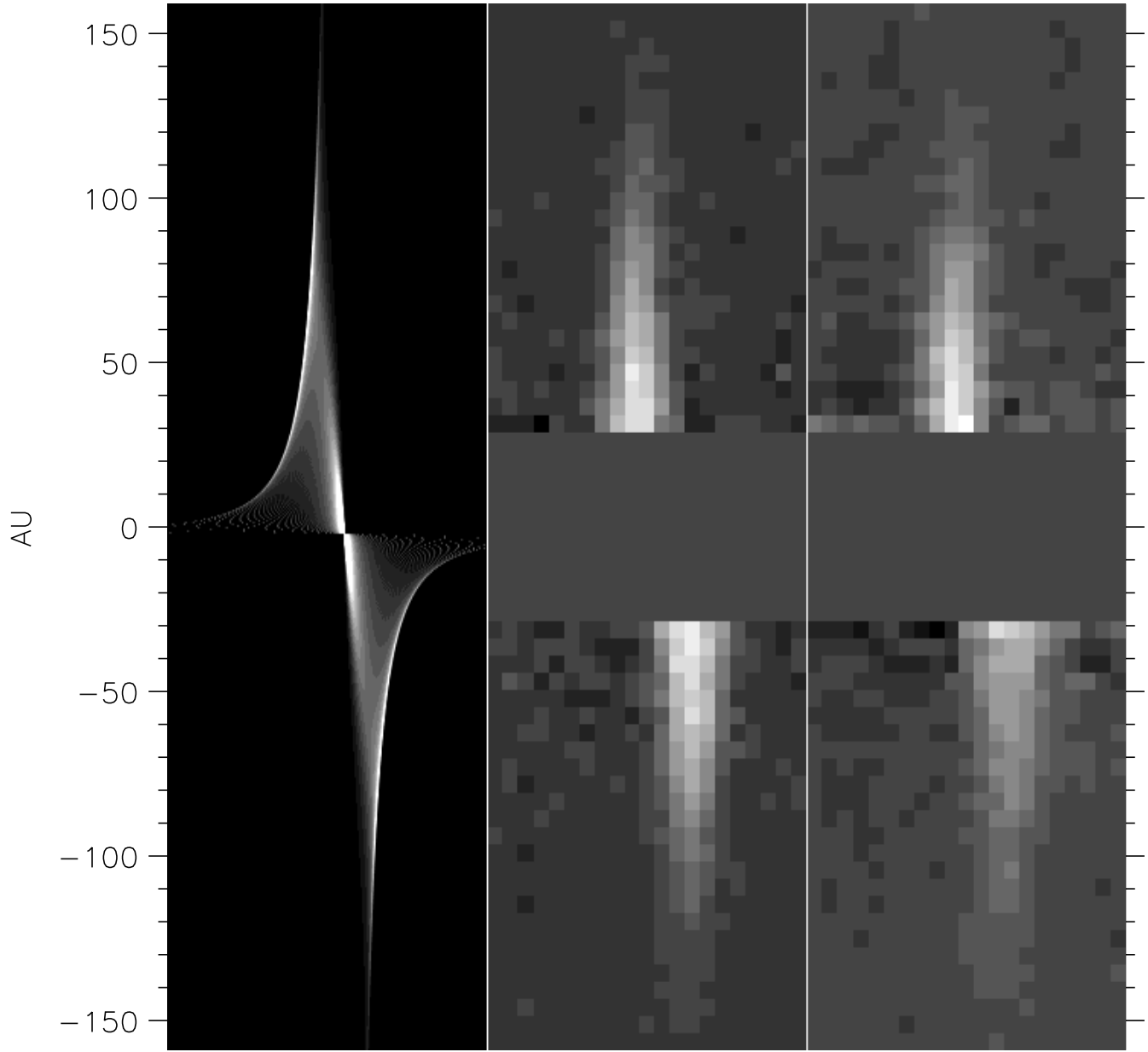}
\caption{Comparison of the observed \ion{Na}{1} D2 line with the results of
theoretical model calculations. To the left, the model is displayed at
spatial and spectral resolutions 10 times better than the observed. In
the middle we show the model degraded to the quality of the observations
(including noise). As is seen, it compares well with the observations
(to the right).\label{fig5}}
\end{figure}

Our observations are consistent with circular Keplerian rotation of the gas
(see Fig.~\ref{fig5}), were we have taken the
measured seeing and spectral instrumental profile (using telluric
absorption lines) into account. The best-fit mass, in a $\chi^2$-sense, 
equals ($1.40\pm0.05$)\,\msun.  This model mass is less than that deduced
from the stellar position in
the HR-diagram, viz. 1.75\,\msun\ (e.g. \citet{cri97}). It is possible 
that this difference is due to radiation pressure (see Sect.\,3.3).

\subsection{The light budget}

The origin of the ``stable gas'' absorption has been extensively discussed
and there are arguments for a location very close to the star \citep{lag98}.
If, on the other hand, the sodium emission far out in the
disk is caused by resonance scattering of stellar radiation then, at
least partly, the ``stable gas'' absorption must originate throughout the
disk along the line of sight. We measure an \ion{Na}{1}\,D2 equivalent width in
absorption of 9.4\,m\AA, implying a column density of
$N({\rm Na\,I}) = 7\cdot 10^{10}\,{\rm cm}^{-2}$, in agreement with previous
observations by \citet{vid86}. In emission, a total equivalent width of
0.72\,m\AA\ is obtained. This latter number is of course a lower limit as it does
not include any line emission originating outside the one arcsecond wide
slit of the spectrograph. It means that the disk, as seen from the star,
must occupy at least a latitudinal angle of \adegdot{8}{8}. We start to
detect the line emission at a distance of 30\,AU and at this distance the
required thickness of the disk would be at least 4.6\,AU. As the slit
width covers 19\,AU these estimates do not lead to any contradiction
regarding the light budget under the assumption of resonance scattering.
On the other hand, there is not much margin for the proposed dense \ion{H}{1}
ring \citep{lag98} to contribute to the stable \ion{Na}{1} absorption lines.

\subsection{The radiation pressure}

How, then, can we understand why the radiation pressure does not quickly
accelerate the \ion{Na}{1} outwards in the same sense as it does, for instance,
in comets (cf. \citet{cre97})? We first note that the momentum transfer caused
by the resonance scattering vastly exceeds the gravity. Assuming a
stellar mass of 1.75\,\msun\ we find that the force caused by the radiation
pressure exceeds the gravitational force by a factor of 300. A possible
explanation why we do not observe a radial velocity component is the
presence of a relatively dense gas component. However, this would require
very large amounts of gas far out in the disk. As an example we consider
the distance of 100\,AU. Our observations exclude a radial velocity component exceeding
5\,\kms\ and assuming that the main gas component is atomic hydrogen we find
that the gas density must be at least $10^5\,{\rm cm}^{-3}$. As the thickness (or
scale height) of the gas component of the disk is unknown, it is hard to
judge if this minimum density is in conflict with other observations.

Even though we cannot at present rule out this explanation for the
balancing of the radiation pressure, we propose an alternative: Due to
the low ionisation potential of \ion{Na}{1} (5.1\,eV), one would expect the
stellar UV radiation to provide a high degree of ionisation. As \ion{Na}{2}
lacks strong transitions within the range of the stellar spectrum,
the radiation pressure on the ionised sodium gas does not
counterweigh the gravitation. Thus, if the gas density is low, a sodium
atom will quickly take up speed outwards. But it will stay neutral just
for a short time and then, as singly ionised, it will have a long time to
adapt to the motion of the main gas component.

\begin{figure}
\plotone{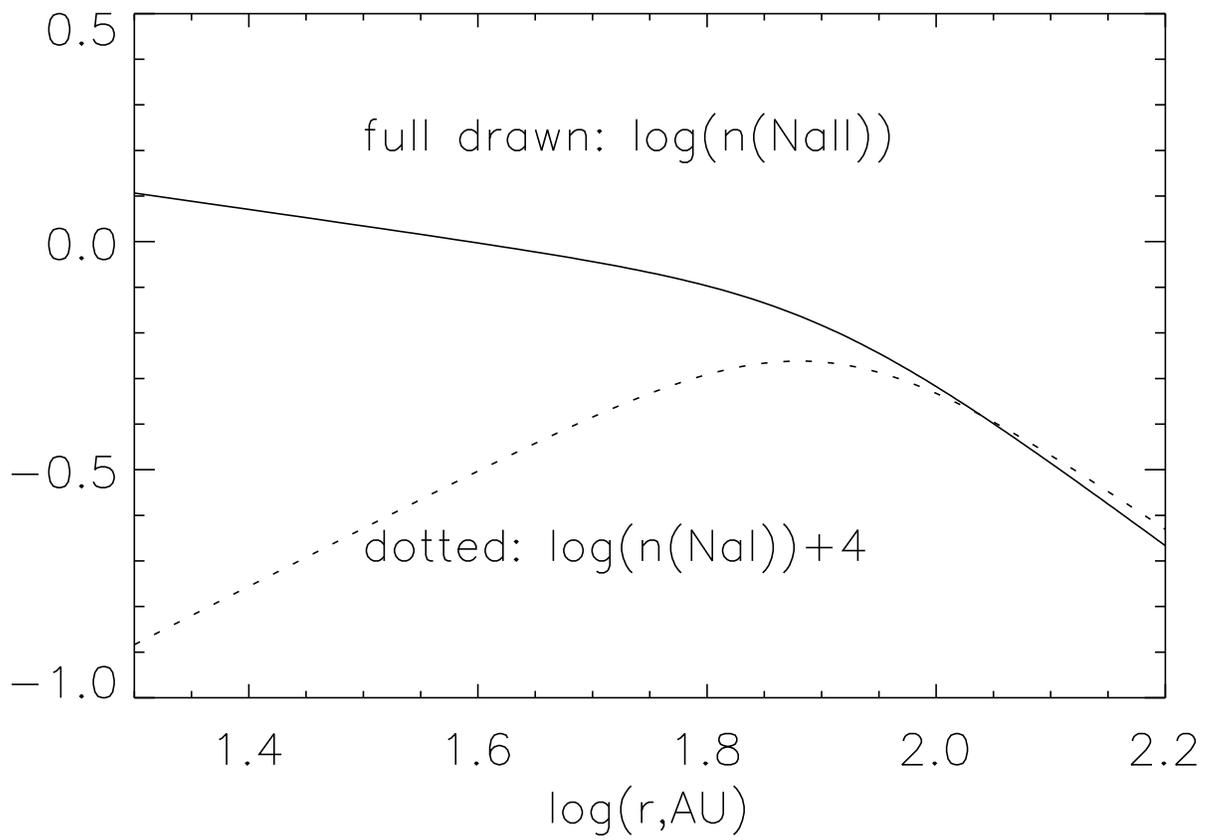}
\caption{The derived number density (${\rm cm}^{-3}$) for \ion{Na}{1} and \ion{Na}{2}.
Only a small fraction ($< 10^{-4}$) of the sodium is neutral.\label{fig6}}
\end{figure}

To quantitatively test
this idea we estimate the degree of ionisation. As we can exclude
higher ionisation stages (the ionisation potential is 47\,eV for \ion{Na}{2}) the
equation of ionisation equilibrium reads
\begin{displaymath}
\frac{n{\rm (Na\,II)}}{n{\rm (Na\,I)}}
 = \frac{\omega (r)\, \Gamma\!_0 }{\alpha_{{\rm tot}}\left [T(r) \right ]\,n_{\rm e}(r)}
\end{displaymath}
where $\omega$ is the dilution factor of the radiation density at the
distance $r$ from the star compared to that at the stellar surface, $\Gamma\!_0$ is the
ionisation rate at the stellar surface, $\alpha_{{\rm tot}}$ is the total
recombination coefficient and $n_{\rm e}$ the electron density. Using a model
atmosphere from \citet{all00} and cross sections from \citet{cun93}
we estimate $\Gamma\!_0= 205\,{\rm s}^{-1}$. Assuming 100\,K as a typical
electron temperature in the disk we get $\alpha_{{\rm tot}}=3.85\cdot
10^{-12}\,{\rm cm}^3{\rm s}^{-1}$ \citep{ver96}.
The electron density remains to be estimated. The stellar far-UV continuum
does not suffice to ionise H, He (if at all present), C, N and O.
Assuming solar abundances \citep{hol95} for elements with 
ionisation potentials $\le 8.3$\,eV (longward of 1500\,\AA) we estimate $n_{\rm e} \sim 
50\,n({\rm Na\,II})$.
If we, finally, assume a constant thickness of the disk (the half power
width of the dust disk is in fact close to constant from 30\,AU and
outwards, see \citet{hea00}) we can use the radial distribution
derived for \ion{Na}{1} in combination with the column density to
derive the degree of ionisation as a function of the
distance to the star. We find that sodium is indeed highly ionised with
$n{\rm (Na\,I)}/n{\rm (Na)} < 10^{-4}$ throughout the disk. In Fig.~\ref{fig6},
we show the radial distribution of $n$(\ion{Na}{1}) and $n$(\ion{Na}{2}).
These analytical results have been confirmed by self-consistent photoionisation
computations, which produce the observed line fluxes. The details will be
given in a forthcoming article.

We conclude that the radiation pressure on sodium, averaged over time,
would be small. We must, however, also estimate what speed a neutral
atom would typically achieve until it becomes ionised. Both the UV
continuum and the sodium D lines are optically thin throughout the disk
and thus both ionisation rate and the resonance scattering rate roughly
scale as the inverse square of the distance to the star. Therefore, the 
number of
scattered photons per period of neutral state is constant and we find
that typically $1.5\cdot 10^4$ scattering events will occur until an atom
becomes ionised. The scattered radiation is essentially isotropic and the
transferred momentum would cause an outward velocity of merely 0.4\,\kms.
After that, there will be a long period of time during which the sodium ion
will interact with the main gas components and conform to the general
velocity pattern, be it basically circular Keplerian rotation or not.
This also means that the radiation pressure may cause a slow, stepwise
motion of each sodium atom outwards resulting in a net flow from the
inner to the outer parts of the disk.

\subsection{The amount of atomic gas in the disk}

 From the radial distribution of \ion{Na}{1} we were able to derive the radial
distribution of \ion{Na}{2} (Fig.~\ref{fig6}). The number density is typically
$n$(Na)=$1\,{\rm cm}^{-3}$ and the column density from 30 to 140\,AU\ is
$N$(Na)=$10^{15}\,{\rm cm}^{-2}$. If the Na/H
ratio were solar, this would indicate a hydrogen column density of $N$(H) = $5\cdot
10^{20}\,{\rm cm}^{-2}$. Further discussion of these aspects will be deferred to a
forthcoming paper.

\section{Conclusions}

We have observed \ion{Na}{1} D1 and D2 emission along the disk of $\beta$\,Pic from
a projected distance of 30 to 140\,AU. The velocity pattern mimics
circular Keplerian rotation, but the deduced mass is somewhat lower than
that expected for an A5V star (1.4 compared to 1.75\,\msun). This
difference is probably due to radiation pressure that to some extent
counterbalances gravity. We find that the sodium gas is mainly
ionised, with \ion{Na}{2}/\ion{Na}{1} around $10^4$ and for this reason,
the radiation pressure does not accelerate the sodium gas component to high
velocities. This is simply because the number of scattering events during
a neutral period of a sodium atom only would suffice to give an outward
velocity of 0.4\,\kms\ and then, during the typically $10^4$ times longer
period of singly ionised state, the radiation pressure is negligible. So,
there is plenty of time for the sodium gas to conform, through gas friction,
to the motion of the main gas component (i.e. H, O, C and N)
that has only a weak direct interaction with the radiation field.

It is, finally, clear that the observations presented in the present
{\it Letter} only mark a starting point of spectroscopic investigations as
both the spectral and the spatial resolutions as well as the straylight
level can be significantly improved upon. In addition, other lines like
the \ion{Ca}{2} H\,\&\,K doublet and a number of resonance lines in the UV,
accessible from e.g. the HST, will probably be detected in the disk.

\clearpage

\end{document}